\documentclass[twocolumn,english,aps,prd,reprint,floatfix,notitlepage,footinbib,preprintnumbers,superscriptaddress,longbibliography]{revtex4-1}
\pdfoutput=1
\usepackage{lmodern}

\usepackage[T1]{fontenc}
\usepackage[latin9]{inputenc}
\usepackage{geometry}
\usepackage{subfigure,lmodern, amsmath,amssymb, graphicx, pifont, adjustbox, bm, xcolor}
\usepackage{amsfonts}
\usepackage{enumitem}
\usepackage{comment}
\usepackage{mathtools}
\usepackage{float}
\usepackage{slashed}
\usepackage{ragged2e}
\usepackage{array}
\usepackage{bbm}

\usepackage{nameref}

\usepackage{hhline}

\makeatletter\g@addto@macro\bfseries{\boldmath}\makeatother

\makeatletter\newcommand{\labeltext}[2]{%
  \def\@currentlabel{#1}%
  \label{#2}%
}
\makeatother

\usepackage{stackengine}
\usepackage{esint}
\usepackage[unicode=true,pdfusetitle,
 bookmarks=true,bookmarksnumbered=false,bookmarksopen=false,
 breaklinks=false,pdfborder={0 0 1},backref=false,colorlinks=true]
 {hyperref}
\hypersetup{
 pdfauthor={Francesco Calisto, Clifford Cheung, Grant N. Remmen, Francesco Sciotti, Michele Tarquini},
 citecolor=black,linkcolor=black,urlcolor=black}

\newcommand{\appendixref}[1]{\hyperref[#1]{appendix~\ref{#1}}}
\def\equationautorefname~#1\null{eq.\,(#1)\null}
\usepackage{breakurl}
\usepackage[hang,flushmargin]{footmisc} 
\allowdisplaybreaks
\makeatletter

\usepackage{etoolbox}
\apptocmd{\thebibliography}{\justifying\setlength{\leftskip}{7.4mm}}{}{} 
 
 \usepackage{relsize}
\usepackage{babel}

\makeatletter
\def\simgt{\mathrel{\lower2.5pt\vbox{\lineskip=0pt\baselineskip=0pt
           \hbox{$>$}\hbox{$\sim$}}}}
\def\simlt{\mathrel{\lower2.5pt\vbox{\lineskip=0pt\baselineskip=0pt
           \hbox{$<$}\hbox{$\sim$}}}}
\makeatother

\usepackage{changepage}

\newcommand{\be}{\begin{equation}}
\newcommand{\ee}{\end{equation}}
\newcommand{\bea}{\begin{eqnarray}}
\newcommand{\eea}{\end{eqnarray}}

\newcommand{\Eq}[1]{Eq.~(\ref{#1})}
\newcommand{\Eqs}[2]{Eqs.~(\ref{#1}) and (\ref{#2})}

\newcommand{\eq}[2]{\be\begin{aligned}#1 \label{#2}\end{aligned}\ee}
\newcommand{\mysec}[1]{\noindent {\bf #1.}---}

\newcolumntype{P}[1]{>{\centering\arraybackslash}p{#1}}

\usepackage{fix-cm}

\definecolor{dartmouthgreen}{rgb}{0.05, 0.5, 0.06}

\begin{document}

\preprint{CALT-TH 2026-015}

\title{The Equivalence Principle at High Energies Completes the Spectrum}

\author{Francesco Calisto}
\affiliation{Walter Burke Institute for Theoretical Physics and
Leinweber Forum for Theoretical Physics, California Institute of Technology, Pasadena, CA 91125, USA}
\author{Clifford Cheung}
\affiliation{Walter Burke Institute for Theoretical Physics and
Leinweber Forum for Theoretical Physics, California Institute of Technology, Pasadena, CA 91125, USA}
\author{Grant N.~Remmen}
\affiliation{\scalebox{1}{Center for Cosmology and Particle Physics, Department of Physics, New York University, New York, NY 10003, USA}}    
\author{Francesco Sciotti}
\affiliation{IFAE and BIST, Universitat Aut\`onoma de Barcelona, 08193 Bellaterra, Barcelona, Spain}
\author{Michele Tarquini}
\affiliation{Walter Burke Institute for Theoretical Physics and
Leinweber Forum for Theoretical Physics, California Institute of Technology, Pasadena, CA 91125, USA}

\begin{abstract}

\noindent We prove a version of the completeness hypothesis that follows from the coexistence of symmetry and gravity:  tree-level gravitational scattering mandates single-particle states in all possible irreducible representations of the symmetry group constructible from a single seed charge.  Our main assumption is that the leading high-energy behavior of scattering is universal irrespective of charge, thus satisfying the equivalence principle.  Curiously, we discover that these newly-deduced states contribute democratically---that is, with equal interaction strengths---to scattering.

\end{abstract}

\maketitle

\mysec{Introduction}What are the universal features of any consistent theory of quantum gravity?  By its very nature, this question transcends string theory or any alternatives.   Indeed, this challenge pertains to the {\it full space} of possible ultraviolet completions of general relativity, whatever that may be.

Along these lines, the swampland program~\cite{Vafa:2005ui} has identified a litany of seemingly universal properties of all known string theoretic constructions.  These patterns have been articulated as broad conjectures dictating the fate of symmetry in any consistent theory of quantum gravity: exact global symmetries are forbidden~\cite{Banks:1988yz,Harlow:2018tng}, gravity is the weakest force~\cite{Arkani-Hamed:2006emk}, and any charged state that is not forbidden is mandatory~\cite{Polchinski:2003bq,Banks:2010zn}.
The latter is known as the completeness hypothesis.

In parallel with these top-down efforts, the modern scattering amplitudes program has developed bootstrap methods that stringently constrain the space of consistent theories from the bottom up.
Axiomatic principles like locality, causality, and unitarity~\cite{Adams:2006sv} have enjoyed great success bounding a variety of effective field theories (EFTs), even finding purchase in placing bounds on quantum gravity and providing theory-independent justifications for conjectures from the swampland~\cite{completeness,Hillman:2024ouy,Arkani-Hamed:2021ajd,Cheung:2018cwt,Cheung:2019cwi,Pham:1985cr,Ananthanarayan:1994hf,Pennington:1994kc,Fernandez:2022kzi,Jenkins:2006ia,Dvali:2012zc,Arkani-Hamed:2020blm,Caron-Huot:2020cmc,EliasMiro:2022xaa,EliasMiro:2026kww,Caron-Huot:2021rmr,Nicolis:2009qm,Tolley:2020gtv,Bellazzini:2020cot,Bellazzini:2025shd,Bellazzini:2015cra,Cheung:2016wjt,Camanho:2014apa,Gruzinov:2006ie,Arkani-Hamed:2021ajd,Cheung:2018cwt,Cheung:2019cwi,Cheung:2014ega,Bellazzini:2019xts,Andriolo:2020lul,Caron-Huot:2022ugt,Caron-Huot:2022jli,Cheung:2016yqr,deRham:2017xox,Bellazzini:2023nqj,Camanho:2016opx,Bern:2021ppb,Berman:2023jys,Freytsis:2022aho,Green:2023ids,Baumann:2019ghk,Remmen:2019cyz,Remmen:2020vts,Remmen:2020uze,Bellazzini:2016xrt,Remmen:2022orj,Remmen:2024hry,ATLAS:2026wew,Bi:2019phv,Zhang:2018shp,Low:2009di,Englert:2019zmt,ZZ,YZZ,Trott,HYZZ,Chang:2026ztn,Chala:2023xjy,Berman:2024wyt,Remmen:2021zmc,Bachu:2022gof,Huang:2022mdb,Cheung:2022mkw,Cheung:2023adk,Cheung:2023uwn,Haring:2023zwu,Haring:2024wyz,Arkani-Hamed:2022gsa,Bhardwaj:2024klc,Aoude:2024xpx,Cheung:2023hkq,Low:2024hvn,Kowalska:2025qmf,Bresciani:2025toe,Chandrasekaran:2018qmx,SFAN,Guerrieri:2021ivu,Albert:2024yap,Berman:2024eid,Berman:2025owb,Elvang:2026pmc,Cheung:2024uhn,Cheung:2024obl,Arkani-Hamed:2023jwn,Cheung:2025nhw,Basile:2026gnd,Eckner:2024ggx,Eckner:2024pqt,Bucciotti:2025dnh,Eckner:2025kve,Caron-Huot:2016icg,Albert:2022oes,Albert:2023jtd,Albert:2023seb,Ma:2023vgc,Dong:2024omo,Bellazzini:2025bay,Huang:2025icl,Beadle:2025cdx,Beadle:2024hqg,Dong:2025dpy,Pasiecznik:2025eqc,Caron-Huot:2024lbf}.

In this paper we present a simple argument that rigorously proves a version of the completeness hypothesis for gravitational EFTs.   Our principal assumptions  are that there is
$i)$ an exact symmetry group $G$, either discrete or continuous, abelian or nonabelian, gauged or nongauged, $ii)$ a spectrum including at least one particle charged under $G$, and  $iii)$ a weakly coupled ultraviolet completion of gravity in which the leading high-energy behavior satisfies the equivalence principle.    Physically, this last condition encodes the reasonable property that the high-energy behavior of gravitationally interacting particles is universal irrespective of charge. Mathematically, it is the property that the leading Regge trajectory is a singlet under $G$. 
Note that our assumptions are satisfied by string amplitudes, but they apply more broadly.  

Under these assumptions, we use analytic dispersion relations for scattering amplitudes to rigorously derive {\it representation completeness}: all irreducible representations obtained from the tensor products of a seed charged state must appear as single-particle states in the spectrum.   Building on the program first initiated in Ref.~\cite{Hillman:2024ouy}, the present work establishes a much stronger conclusion than an earlier study~\cite{completeness}, which used a distinct and more complicated analysis to derive completeness for abelian subgroups of certain continuous nonabelian symmetries $G$ of sufficiently high rank.

An intriguing corollary of the current analysis is {\it representation democracy}:  the particles whose existence we deduce contribute with squared couplings that are completely independent of their representation after averaging over all quantum numbers distinct from $G$.   Said another way, particles contribute to scattering democratically with respect to their representation.
Since these states are the asymptotically high-energy excitations in the spectrum, the correspondence principle \cite{Horowitz:1996nw} suggests that this property may be a manifestation of equipartition of states into charge sectors in a black hole.

\medskip

\mysec{Amplitude}Consider a tree-level four-point amplitude for particles charged under an exact symmetry $G$ in a gravitational EFT.  In principle, this symmetry may be global or gauged.  We will not consider the case where it is explicitly broken.
Without loss of generality, the low-energy amplitude of massless particles is
\eq{
A(z,t) =& -\frac{8\pi G_N z^2}{t}\mathbbm{1} +\frac{8\pi \alpha z}{t} \mathbbm{T}+ \sum_{k=0}^\infty c_k(t) z^k ,
}{A_def}
where the center-of-mass energy and momentum transfer are encoded in $z=(s-u)/2$ and $t=-(s+u)$, which are odd and even under crossing the $s$ and $u$ channels.
The $1/t$ singularities encode the tree-level exchange of the graviton and, if present, a gauge boson for the symmetry.  The $c_k(t)$ are regular functions of $t$.     This expansion generalizes straightforwardly to massive scattering.

Since the external states reside in irreducible representations of the symmetry group, the amplitude and all of its terms carry implicit indices.  
In particular, the contribution from graviton exchange in \Eq{A_def} is proportional to the {\it identity element} $\mathbbm{1}$ in charge space, while that of the gauge boson is proportional to a nontrivial tensor $\mathbbm{T}$.  The former leaves the charges of the scattering states unscathed, in accordance with the equivalence principle.

Case in point, for $G=SO(N)$ and external states in the fundamental, we have the explicit index form,
\eq{
\mathbbm{1} = \delta^{i_1 i_4} \delta^{i_2 i_3}.
}{identity_expansion_early}
This tensor describes a neutral exchange in the $t$ channel, so the resulting scattering process has the same charge structure as free propagation.
Our logic will hinge on the fact that $\mathbbm{1}$ can be decomposed into a basis of projectors,
\eq{
\mathbbm{1} =\mathbbm{P}_0^{(s)}+\mathbbm{P}_1^{(s)}+\mathbbm{P}_2^{(s)},
}{idendecomp}
for each $s$-channel irreducible representation,
\begin{equation}\hspace{-2mm}
\begin{aligned}
[\mathbbm{P}_0^{(s)}]^{i_1i_2 i_3 i_4} &=  \frac{1}{N}\delta^{i_1 i_2} \delta^{i_3 i_4}\\
[\mathbbm{P}_1^{(s)}]^{i_1i_2 i_3 i_4} &=  \frac{1}{2}(\delta^{i_1 i_4} \delta^{i_2 i_3}\,{-}\,\delta^{i_1 i_3} \delta^{i_2 i_4})\\
[\mathbbm{P}_2^{(s)}]^{i_1i_2 i_3 i_4} &=  \frac{1}{2}(\delta^{i_1 i_4} \delta^{i_2 i_3}\,{+}\,\delta^{i_1 i_3} \delta^{i_2 i_4})\,{-}\,\frac{1}{N}\delta^{i_1 i_2} \delta^{i_3 i_4}.
\end{aligned}\hspace{-2mm}
\end{equation}
Physically, Eq.~\eqref{idendecomp} implies that the charge structure associated with the graviton pole is a democratic sum over all irreducible representations.

The form of \Eq{identity_expansion_early} is  general and independent of the representations of the external states.  Furthermore, with appropriate sign conventions for the projectors, it applies to both the $s$ and $u$ channel, so 
\eq{
\mathbbm{1} = \sum_\rho \mathbbm{P}_\rho^{(s)} =\sum_\rho \mathbbm{P}_\rho^{(u)},
}{identity_expansion}
where $\rho$ labels all irreducible representations
in the direct sum decomposition of the tensor product of incoming states in that channel~\footnote{Our construction applies extremely broadly, namely to any group $G$ where, given two irreducible representations $\rho_{1,2}$, one can write $\rho_1 \otimes \rho_2$ as a direct sum of irreducible representations of $G$. This commonsense fact is true for the vast majority of groups of relevance to physics, including abelian and nonabelian simple Lie groups, discrete groups that are either finite or abelian, all finite products thereof, and more. Relevant mathematical results include the theorems of Weyl, Maschke, Peter-Weyl, and Plancherel; see Refs.~\cite{fulton1991representation,deitmar2014principles} for a review.}. 
For instance, if the $s$-channel fuses states in distinct representations  $\rho_1$ and $\rho_2$, then $\mathbbm{1} = \delta^{i_1 i_4}\delta^{j_2 j_3}$, where $i_{1},i_{4}=1,\ldots,\dim \rho_1$ and $j_{2},j_{3}=1,\ldots ,\dim \rho_2$.  The simple mathematical fact defined in \Eq{identity_expansion} will undergird all of our results.

\medskip
\mysec{Dispersion Relation}We are now equipped to derive a dispersion relation for the EFT coefficients defined by \Eq{A_def}.  Doing so will require certain assumptions about the Regge behavior of the amplitude.  

The components of the amplitude that are even and odd under $s$- and $u$-channel crossing, known as the pomeron and odderon~\cite{Collins:1977jy,Joynson:1975az}, are
\eq{
A^{(\pm)} (z,t)= \frac12(A(z,t) \pm A(-z,t)).
}{}
Any graviton or gauge boson exchanges are encoded in $A^{(+)}$ and $A^{(-)}$,  respectively.
Recall that at fixed $t<0$, the Regge behavior of the even amplitude falls off as
\eq{
\lim_{z\rightarrow \infty} \frac{A^{(+)}(z,t)}{z^2} &=0.
}{bd2}
As rigorously derived in Ref.~\cite{sasha2024}, the above asymptotics are a universal feature of all gravitational theories. For example, string amplitudes exhibit $A^{(+)} \sim z^{2+t}$, in which case \Eq{bd2} is marginal as $t\rightarrow 0^-$.

A crucial additional assumption of the present work  is that the leading Regge behavior of the amplitude satisfies the equivalence principle.   As we will see, this requirement implies that at  fixed $t<0$, the odd amplitude scales as
\eq{
\lim_{z\rightarrow \infty} \frac{A^{(-)}(z,t)}{z^{2-\delta}} &=0,
}{bd1}
for some constant $\delta>0$ that is independent of $t$.  For string amplitudes,  \Eq{bd1} holds because $A^{(-)} \sim z^{1+t}$.

The appearance of $\delta$ in \Eq{bd1} ensures that the large-$z$ growth of $A^{(-)}$ is strictly subleading to $z^2$ in such a way that is not even marginal as $t\rightarrow 0^-$.    If this condition is violated, then $A^{(-)}$ can scale as $z^2$ or worse as $t\rightarrow 0^-$, thus contributing to the leading Regge scaling of the amplitude. The resulting effect is, by definition, odd under crossing, so it cannot be proportional to the identity in charge space.  In such a case, the high-energy behavior of the amplitude $A = A^{(+)}+A^{(-)}$ depends explicitly on the charges of the external states, in violation of the equivalence principle.  An example of this is $A^{(-)} \sim (s^{2+t}-u^{2+t})/t \sim  i(\arg (z)-\arg(-z)) z^2$, which is odd under crossing but scales as $z^2$.  Though not strictly pathological---pure consistency only requires the equivalence principle in the $t$-channel pole mediating the long-range gravitational force---such high-energy behavior is nevertheless very peculiar.  In any case, our assumption of a high-energy equivalence principle forbids this possibility, so \Eq{bd1} applies.   See the appendix for a discussion of various putative amplitudes that actually violate \Eq{bd1} and the equivalence principle.

For $t<0$, the following contour integral at infinity is zeroed out by the assumption in  \Eq{bd2},
\eq{
\oint_\infty \frac{dz}{z^3} A(z,t)=\oint_\infty \frac{dz}{z^3} A^{(+)}(z,t)=0.
}{bdry}
Deforming the contour of integration inwards to encircle all enclosed poles, we then obtain~\footnote{If there are poles in $z$ induced by the infrared degrees of freedom, for instance from $s$- and $u$-channel exchanges, then our prescription is to choose an initial contour that encloses all of these light particle poles, rather than just the origin.  By Cauchy's theorem, this choice will yield Eq.~\eqref{DR2}, where $n$ starts at the first massive state.}
\eq{
 -\frac{8\pi G_N \mathbbm{1}}{t}+c_2(t) = \sum_n \frac{R^{(+)}(n,t)}{\mu(n,t)^3} ,
}{DR2}
where we have defined $\mu(n,t) = \mu(n) +t/2$ in terms of the spectrum of nonzero squared masses $\mu(n)$.  The amplitude exhibits poles at $s=\mu(n)$ and $u=\mu(n)$~\footnote{We can choose our external states so that their masses satisfy $m_1=m_4$ and $m_2=m_3$. This choice guarantees that the spectrum of squared masses exchanged in the $s$ and $u$ channel are the same, both described by $\mu(n)$. If the external scattered states are massive, then as in Ref.~\cite{Cheung:2016yqr} we should define the  subtraction in Eq.~\eqref{bdry} as
\begin{equation*}
\qquad\;\,\protect\oint_{\infty} \frac{dz}{\left(z+\frac12\protect\sum_{i=1}^4 m_i^2\right)^3}A^{(+)} = 0, 
\end{equation*}
so \Eq{DR2} is unchanged, and similarly for \Eq{c2tilde}.  
}, and without loss of generality,  the ordering of the label $n$ is chosen so that $\mu(n)$ increases monotonically.  We further assume that $\mu(n)$ diverges at large $n$, thus forbidding the possibility of accumulation point spectra~\cite{Coon:1969yw,Maldacena:2022ckr}. 
Here we have also defined the components of the residue that are even and odd under crossing symmetry,
\eq{
R^{(\pm)}(n,t) &=R^{(s)}(n,t)\pm R^{(u)}(n,t)\\
R^{(s)}(n,t) &= \lim_{z\rightarrow+\mu(n,t)}(\mu(n,t)-z)A(z,t)\\
R^{(u)}(n,t) &= \lim_{z\rightarrow-\mu(n,t)}(\mu(n,t)+z)A(z,t).
}{}
where $R^{(s)}(n,t)$ and $R^{(u)}(n,t)$ are the residues at $s=\mu(n)$ and $u=\mu(n)$, respectively.
Note that legs $1$ and $4$ must be in the same irreducible representation, and similarly for $2$ and $3$, so that a graviton can be exchanged in the $t$ channel.
 Locality implies that each of these residues is a polynomial in $t$, describing a finite set of spins exchanged at that mass level. 

Next, we derive a formula for a version of \Eq{DR2} with $R^{(-)}$ instead of $R^{(+)}$,
\eq{
\tilde c_2(t) =  \sum_n \frac{R^{(-)}(n,t)}{\mu(n,t)^3}.
 }{c2tilde}
The power of $\mu(n,t)$ in the denominator is odd rather than even, so $\tilde c_2(t)$ does not correspond to a Wilson coefficient that can be extracted from the amplitude. Nonetheless, it can be computed via a line integral along the imaginary $z$-axis intersecting the origin,
 \eq{
\tilde c_2(t)  &=\frac{1}{\pi i} \int_{-i\infty}^{+i\infty} \frac{dz}{z^3}  \hat A^{(-)} (z,t),
 }{im_contour}
 where $\hat A^{(-)} (z,t) = A^{(-)} (z,t)-  z (\partial_{z'} A^{(-)} (z',t)|_{z'=0})$.  In the small-$z$ expansion, $\hat A^{(-)} = O(z^3)$ is simply $ A^{(-)}  = O(z^1)$ with its $z^1$ component subtracted away.  Crucially, the integrand of \Eq{im_contour} is regular at $z=0$, and the integral converges at infinity due to \Eq{bd1}. We thus deform the contour to enclose all poles of $\hat A^{(-)} (z,t)$  in the right half-plane, where boundary terms at infinity vanish by \Eq{bd1}. Since $\hat A^{(-)} (z,t)$  differs from $A^{(-)} (z,t)$ by an entire function, they share the very same poles at $z= \pm \mu(n,t)$, so
 \eq{
  A^{(-)} (z,t) = \sum_n \frac{z R^{(-)}(n,t)}{\mu(n,t)^2-z^2} +\cdots,
 }{}
 where the ellipses are nonsingular.  Summing the residues in this formula for each $z=+\mu(n,t)$ yields \Eq{c2tilde}.
 
 If the symmetry is gauged, then $A^{(-)} (z,t)$ exhibits a $z/t$ pole that is precisely cancelled in $\hat A^{(-)} (z,t)$.  Consequently, the latter is regular as $t\rightarrow 0^-$.
 This regularity of course persists if the symmetry is global.  Either way, we learn  that $\tilde c_2(t)$ is, crucially, finite as $t\rightarrow 0^-$ \footnote{Mathematically, regularity of $\tilde c_2(t)$ in the  $t\rightarrow 0^-$ limit requires uniform boundedness of \Eq{im_contour}.  For this reason, we assume that there exist $t$-independent constants $c>0$ and $\delta >0$ such that 
  $|\hat A^{(-)} (z,t) | < c |z^{2-\delta}|$ along the full contour of integration in $z$ and on an open set $-\epsilon < t<0$ for some $\epsilon>0$. This statement is a more refined version of the assumption in \Eq{bd1}.  Physically, this assumption says that the odd amplitude---after subtracting its massless $t$-channel pole---falls off sufficiently fast that the equivalence principle remains valid at high energies.
 }.

\medskip
\mysec{Completeness}We are now equipped to derive a remarkable fact about the spectrum of states. 
 Adding and subtracting \Eq{DR2} and \Eq{c2tilde} yields
\eq{
 -\frac{8\pi G_N \mathbbm{1}}{t}+c_2(t)+\tilde c_2(t) &= 2\sum_n \frac{R^{(s)}(n,t)}{\mu(n,t)^3} \\
  -\frac{8\pi G_N \mathbbm{1}}{t}+c_2(t)-\tilde c_2(t) &= 2\sum_n \frac{R^{(u)}(n,t)}{\mu(n,t)^3} ,
}{}
which as $t\rightarrow 0^-$ implies that
\eq{
\sum_n \frac{R^{(s)}(n,t)}{\mu(n,t)^3} \sim \sum_n \frac{R^{(u)}(n,t)}{\mu(n,t)^3} \sim   -\frac{4\pi G_N \mathbbm{1}}{t},
}{diverge}
where we drop all terms that are subleading in $t$.

\begin{figure}[t]
\includegraphics[width=0.9\columnwidth]{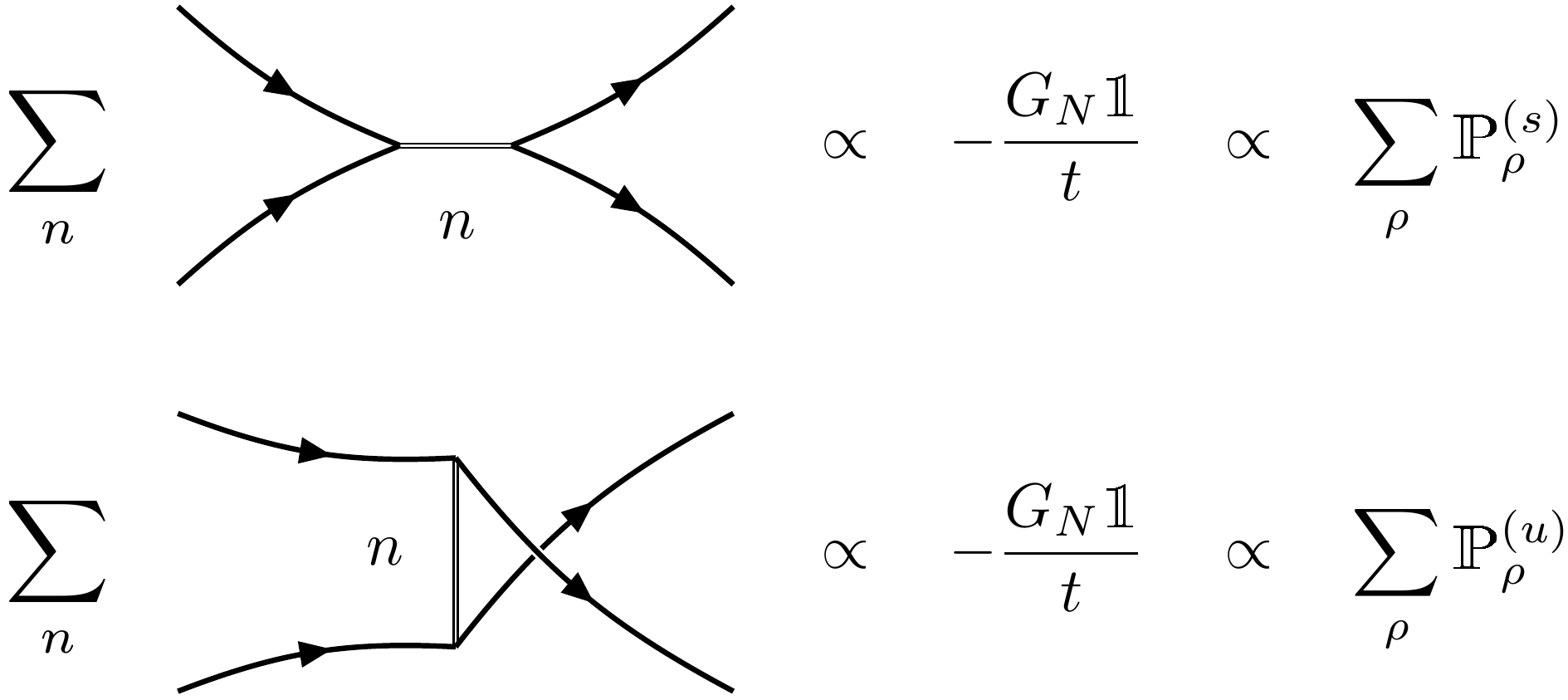}
\caption{According to Eq.~\eqref{diverge}, a certain weighted sum of residues in both the $s$ and $u$ channels must resolve to the identity, thus implying completeness: each representation of the symmetry group must be exhibited by a single-particle state in the spectrum.  Furthermore, the massive states that dictate the Regge behavior are not only complete, but also appear democratically, with interaction strengths that are independent of their representation.}
\label{fig:scattering}
\end{figure}

Equation~\eqref{diverge} implies that the spectrum of irreducible representations of $G$ is complete.   To arrive at this conclusion, we appeal to proof by contradiction: if some irreducible representation is completely absent from all $s$-channel exchanges, then by \Eq{identity_expansion} it is impossible for the aggregate of $s$-channel states to sum to $\mathbbm{1}$, as required by \Eq{diverge}.    Identical logic of course applies to states in the $u$ channel. Thus, as depicted in Fig.~\ref{fig:scattering}, all irreducible representations that arise from the tensor product of the external states fusing in either the $s$ or $u$ channel must appear as single-particle states in the spectrum. 

Following the logic of Ref.~\cite{completeness}, we repeat this process ad infinitum. In particular, we take the $s$-channel states whose existence we have just deduced and then rescatter them as external states,  again assuming Eqs.~\eqref{bd2} and \eqref{bd1}.  Applying the very same dispersion relations as before, we then deduce that all irreducible representations built from these new external states must also be present in the spectrum.
Repeating this process forever, we arrive at the conclusion that every irreducible representation of $G$ generated by tensor products of the original seed must be present in the spectrum.
Since our amplitudes are tree-level into the ultraviolet, each irreducible representation appears as a single-particle state \footnote{Given that an incomplete spectrum is demonstrably inconsistent, it is then natural to ask: At what energy scale do the spectrum-completing states actually reside?  Obviously, if a dispersion relation is only marginally convergent, then one expects states to continue contributing to the dispersion relation up to very high energies.   Quantitatively, given Regge behavior of the form $A \sim s^{\alpha(t)}$ for $\alpha(t)$ small and negative, convergence requires summing states at least up to the scale $ s \gtrsim e^{1/|\alpha(t)|}$.  In all extant works on dispersion relations for scattering, $\alpha(t)$ is taken to be order one or less in units of the gap, in which case one generically expects the dispersion relation to include states that are exponentially heavy relative to that gap~\cite{Haring:2024wyz}.  The same is true for us here.}.
In fact, Eq.~\eqref{diverge} implies that each irreducible representation appears infinitely many times, in an infinite tower of ever more massive single-particle states.

Note that the logic outlined above is vastly more powerful than that of Ref.~\cite{completeness}, which implemented a much more laborious procedure to derive completeness for the abelian charges within certain nonabelian symmetry groups of higher rank.  
The major technical difference between these works is  the high-energy equivalence principle in \Eq{bd1}, which we leverage to derive universality of irreducible representations in both the $s$ {\it and} $u$ channel. In contrast, the argument of Ref.~\cite{completeness} relied on more conservative assumptions about ultraviolet scaling---requiring only $A \lesssim s^2$ and not Eq.~\eqref{bd1}---and employed a technically involved analysis of the root lattice of various symmetry algebras to deduce new states in either the $s$ {\it or} $u$ channel.   As a result, Ref.~\cite{completeness} only established completeness for the abelian charge lattice, while the present work does so for all irreducible representations of any symmetry.

It is worth noting that  $G$ need not be internal, and in fact it can be a symmetry of spacetime.   For example, in the forward limit, since all ingoing and outgoing particles are collinear, their transverse polarizations are unambiguously classified as irreducible representations of the $D$-dimensional massless little group $SO(D-2)$.  
After recasting little group charge as an internal quantum number~\cite{Bellazzini:2016xrt,Remmen:2020uze}, the above logic implies completeness of all irreducible representations of  $SO(D-2)$.
This prediction is borne out by explicit calculation of the spectrum of spinning excitations of the string.  For example, see Ref.~\cite{Hanany:2010da}, which demonstrates little group completeness for the bosonic, heterotic, type~I, and type~II strings, as well as Ref.~\cite{Pesando:2024lqa}, which tabulates the Young tableaux describing the exotic spin states of the bosonic string.
Of course, if the seed particle is a fermion, then we can upgrade to $Spin(D\,{-}\,2)$ to infer completeness among all possible fermionic representations as well.

\medskip
\mysec{Democracy}Let us now consider the Regge limit of the amplitude.  At fixed $t$, we parameterize the behavior of the amplitude at large positive and negative $z$ by
\eq{
\lim_{z\rightarrow + \infty} A&\sim (-z)^{\alpha^{(s)}(t)} \mathbbm{P}^{(s)}_{\infty}\\
\lim_{z\rightarrow - \infty} A&\sim (+z)^{\alpha^{(u)}(t)} \mathbbm{P}^{(u)}_{\infty}.
}{Regge}
Here $[\mathbbm{P}^{(s)}_{\infty}]^{i_1j_2 j_3 i_4}$ and  $[\mathbbm{P}^{(u)}_{\infty}]^{i_1j_2 j_3 i_4}$ are unspecified tensors encoding the symmetry structure of the Regge behavior in the $s$ and $u$ channels.  Other tensor structures can appear, but only one will dominate the Regge limit.

Equation~\eqref{Regge} exhibits branch cuts along the positive and negative real $z$-axis.   As is well known, these  nonanalyticities encode the infinite tower of simple poles in the $s$ and $u$ channel~\cite{SFAN}.  Said another way, the branch cuts are a smeared approximation of the simple poles, averaged over some resolution in energy squared that we define to be $\Delta z$, or equivalently $\Delta n$ in level spacing.  By definition, we take $\Delta z$ and $\Delta n$ to be the spacing at which the Regge formula in \Eq{Regge} is a valid approximation of the actual amplitude.

Next, we compute the weighted integral of the $s$-channel discontinuity of the amplitude on an interval and equate it to the corresponding sum over residues,
\eq{
 \sum_{n'=n}^{n + \Delta n} \frac{R^{(s)}(n',t)}{\mu(n')^3} \sim \frac{1}{2\pi i}\int\limits_{z}^{z+\Delta z} \frac{dz'}{{z'}^3} \,  {\rm disc}(A) \propto \mathbbm{P}^{(s)}_{\infty},
}{}
and likewise for the $u$ channel.
Assuming that the variation in $\mu(n)$ over $\Delta n$ is much smaller than $\mu(n)$ itself, it is then natural to define the averaged residues,
\eq{
\overline{R}^{(s)}(n,t)&= \frac{1}{\Delta n}\sum_{n'=n}^{n + \Delta n}  R^{(s)}(n',t),
}{}
and similarly for $u$.
Comparing to \Eq{diverge}, we then deduce that $\mathbbm{P}^{(s)}_\infty =\mathbbm{P}^{(u)}_\infty= \mathbbm{1}$, and we arrive at the condition of representation democracy,
\eq{
\overline{R}^{(s)}(n,t) \sim \overline{R}^{(u)}(n,t) \sim \mathbbm{1},
}{yay_democracy}
where according to \Eq{identity_expansion} the right-hand side is a democratic superposition of irreducible representations.

By definition, the residue $R^{(s)}(n,t)$ is a product of three-point amplitudes, with the exchanged state summed over all quantum numbers at that level $n$.   Then $\overline{R}^{(s)}(n,t)$ averages this quantity over a range of levels.  The result in \Eq{yay_democracy} thus  implies that this average squared coupling is equally distributed among representations, so all representations contribute democratically to the scattering process near level $n$. 

According to the correspondence principle~\cite{Horowitz:1996nw}, sufficiently high-level resonances of the string should have a dual description as black holes.  This transition is expected when the Schwarzschild radius of a given string mode exceeds the string length.  Blithely interpreting \Eq{yay_democracy} from this perspective, it is tempting to conclude that the black hole states produced in scattering exhibit a natural equipartition among all possible charged states.  
We leave the question of whether this connection can be made rigorous to future work.

\medskip

\mysec{Heterotic String Amplitudes}It is illuminating to verify our claims for a well-behaved ultraviolet completion containing gravity and gauge theory: the heterotic string.
The tree-level amplitude for the two-to-two scattering of gauge bosons in heterotic string theory is
 \eq{
\begin{aligned}
A =& -K\frac{\Gamma(-s/2)\Gamma(-t/2)\Gamma(-u/2)}{\Gamma(s/2)\Gamma(t/2)\Gamma(u/2)} \times \\
&\times \bigg[\left(\frac{{\rm tr}(T_1 T_2){\rm tr}(T_3 T_4)}{16s(1+s/2)} + \frac{15{\rm tr}(T_1 T_2 T_3 T_4)}{st}\right)  \\
& + \left(\begin{array}{c} 1234\,{\rightarrow} \,1423 \\ s,t,u \,{\rightarrow }\, t,u,s \end{array}\right ) {+} \left(\begin{array}{c} 1234\,{\rightarrow} \,1342 \\ s,t,u\,{\rightarrow} \,u,s,t \end{array}\right ) \bigg],
\end{aligned}
}{}
where the external states are adjoints of  $G=SO(32)$ or $SO(16)\otimes SO(16) \subset E_8 \otimes E_8$~\cite{Gross:1985rr}. We note the distinctive linear spectrum of squared masses $\mu(n)\propto n$, as evident from the gamma functions.
Here the kinematic invariant $K$ encodes the external polarizations and scales as $p^4$ in the momenta.
At low energies, this amplitude describes graviton and gauge boson exchanges,
\begin{equation}
\hspace{-2mm}\begin{aligned}
&A = K \bigg[\left(\frac{{\rm tr}(T_1 T_2){\rm tr}(T_3 T_4)}{16s} + \frac{15\,{\rm tr}(T_1 T_2 T_3 T_4)}{st}\right) \\&+ \left(\begin{array}{c} 1234\,{\rightarrow} \,1423 \\ s,t,u\,{\rightarrow }\,t,u,s \end{array}\right ) {+} \left(\begin{array}{c} 1234\,{\rightarrow} \,1342 \\ s,t,u\,{\rightarrow}\, u,s,t \end{array}\right )\bigg] + \cdots.
\end{aligned}
\end{equation}
It is straightforward to verify that the Regge behavior of this amplitude at large $s$ and fixed $t$ is
\be 
\hspace{-2mm}\begin{aligned}
\lim_{s\rightarrow \infty} \! A^{(+)} &= s^{2{+}t}\frac{{\cal E}\,{\rm tr}(T_1 T_4){\rm tr}(T_2 T_3)}{16(1+t/2)}\\
\lim_{s\rightarrow \infty} \! A^{(-)} &= 15 s^{1{+}t}{\cal E}\,[{\rm tr}(T_1 T_2 T_3 T_4){-}{\rm tr}(T_1 T_4 T_2 T_3)] \\ 
{\cal E}& =-\frac{\pi (\epsilon_1 \epsilon_4)(\epsilon_2 \epsilon_3)[\cot(\pi s/2) {+} \cot(\pi t/2)]}{2^{t-1}\Gamma(1+t/2)^2},
\end{aligned}
\ee
taking all polarizations transverse to the momenta.
The residues of the heterotic amplitude at large $n$ on the poles at $s=2n$ or $u=2n$ are given by 
\eq{\hspace{-2mm}
\begin{aligned}
&R(n,t) \sim \frac{(\epsilon_1 \epsilon_4)(\epsilon_2 \epsilon_3){\rm tr}(T_1 T_4){\rm tr}(T_2 T_3)}{\Gamma(1+t/2)\Gamma(2+t/2)}n^{2+t}  \\& {\pm} \frac{120(\epsilon_1 \epsilon_4)(\epsilon_2 \epsilon_3) [{\rm tr}(T_1 T_2 T_3 T_4){-}{\rm tr}(T_1 T_4 T_2 T_3)]}{\Gamma(1+t/2)^2} n^{1{+}t},
\end{aligned}\hspace{-2mm}
}{}
where $+$ obtains for $s$ and $-$ for $u$. From both the trace and polarization structure of the $n^{2+t}$ factor, we see that the leading Regge behavior encodes the identity $\mathbbm{1}$ in both the color and polarization indices, so our predictions of representation completeness and democracy are both borne out by the heterotic string.
From here we see that the $1/t$ graviton and gauge boson poles at $O(s^2)$ and $O(s)$ arise from $\sum_n n^{-1+t} = -1/t + \cdots$ once $R(n,t)$ is weighted with $1/\mu(n)^3$ or $1/\mu(n)^2$, respectively.

\medskip

\mysec{Discussion}We have argued that a version of the completeness hypothesis follows from certain well-motivated features of gravitational scattering.  The cornerstone of our argument is the assumption that the equivalence principle---which mandates universality irrespective of the detailed properties of gravitationally interacting bodies---is robust in the regime of high-energy scattering, as encoded mathematically in \Eqs{bd2}{bd1}.  From this input we deduce that for any weakly coupled ultraviolet completion of gravity exhibiting a global or gauge symmetry $G$, the spectrum must contain all possible irreducible representations generated by tensor products of a given seed charge.
A remarkable corollary of our analysis is the phenomenon of representation democracy: not only is the spectrum complete, but all representations of $G$ appear in high-energy scattering with squared couplings of precisely the same size.

Since the present work has focused exclusively on tree-level amplitudes, it is natural to wonder if adding loop-level amplitudes will modify our conclusions.
In string theory, for example, the perturbative loop corrections to the Regge behavior are quite subtle~\cite{mizera2025,Dudas:1999gz,Dimopoulos:2001qe}.  Fortunately, these concerns are immaterial: in any perturbative ultraviolet completion where the unitarizing states enter at tree level, loop corrections are by definition subleading.  If one-loop corrections are as important as tree, then so too are the contributions at two-loop order, and so on.  Within such perturbatively calculable theories, we have established completeness of the spectrum.

The results of this paper contribute to a growing body of evidence that the broad features of string theory might actually follow directly from well-motivated physical conditions on scattering~\cite{Caron-Huot:2016icg,Cheung:2024uhn,Cheung:2024obl,Elvang:2026pmc,Berman:2024eid,Berman:2025owb,Basile:2026gnd,Arkani-Hamed:2023jwn,Cheung:2025nhw,Albert:2024yap,Guerrieri:2021ivu,Arkani-Hamed:2021ajd,SFAN,Hillman:2024ouy,Eckner:2024ggx,Eckner:2024pqt,Bucciotti:2025dnh,Eckner:2025kve,completeness}.  For example, the seminal work of Ref.~\cite{Caron-Huot:2016icg} argued for the asymptotic uniqueness of strings from unitarity and crossing, although an author of that paper later demonstrated~\cite{Haring:2023zwu} explicit counterexamples to their technical assumptions.  In the past year there has been a resurgence of interest in this topic, uniquely bootstrapping strings from alternative assumptions, such as ultrasoftness~\cite{SFAN} or supersymmetry~\cite{Elvang:2026pmc}.
These developments all suggest that there is much promise in characterizing quantum gravity---its states, dynamics, and potentially other swampland conjectures---using the tools of scattering amplitudes.

\medskip

\noindent {\it Acknowledgments:} 
We thank Simon Caron-Huot, Aaron Hillman, and Yu-tin Huang for discussions.
F.C., C.C.,~and M.T.~are supported by the Department of Energy (Grant No.~DE-SC0011632), the Walter Burke Institute for Theoretical Physics, and the Leinweber Forum for Theoretical Physics. 
G.N.R. is supported by the James Arthur Postdoctoral Fellowship at New York University. F.S.~is
supported by research grants 2021-SGR-00649, PID2023-146686NB-C31, and funding from the European Union NextGenerationEU (PRTR-C17.I1).

\bibliographystyle{utphys-modified}
\bibliography{democracy}

\medskip

%\clearpage

\appendix

\mysec{Appendix: High-Energy Behavior}\label{app:X}It is natural to ask whether democracy holds for the open type~I superstring, which has a massless gauge boson but does not exchange a graviton in the planar sector. To investigate this situation, let us replace the ultraviolet scaling assumptions in \Eqs{bd2}{bd1} with lower powers in $s$, $\lim_{s\rightarrow \infty} A^{(-)}/s =0$ and $\lim_{s\rightarrow \infty} A^{(+)} =0$ for $t<0$, since the high-energy scattering of massless gauge bosons in open superstring theory goes like $A\sim A^{(-)} \sim s^{1+t}$.

Following a derivation analogous to the gravity case, but with these different ultraviolet conditions, we would find that the averaged residues asymptotically satisfy
\eq{
\lim_{n\rightarrow \infty}\overline{R}^{(s)}(n,0) \sim -\lim_{n\rightarrow \infty} \overline{R}^{(u)}(n,0) \propto  \mathbbm{P}_{\rm adj}^{(t)},
}{asymptotic2}
where the color structure $\mathbbm{P}_{\rm adj}^{(t)}$ is the projector onto the adjoint state in the $t$ channel, $\propto f^{a_1 a_4 b}f^{a_2 a_3 b}$. 

Writing $\mathbbm{P}_{\rm adj}^{(t)}$ in terms of projectors in the $s$ or $u$ channel when the external particles are gauge bosons does not yield democracy, since it is not $\propto\mathbbm{1}$, but it would still imply the existence of more representations than are present in the actual superstring spectrum. 
At tree level, the open type~I superstring contains only adjoints of the $SO(32)$ gauge group, with each end of the string carrying only a single color index~\cite{Polchinski,BBS}.
This apparent paradox is reconciled by the fact that, for the superstring amplitude, the assumption $\lim_{s\rightarrow \infty} A^{(+)} =0$ for $t<0$---which would have implied Eq.~\eqref{asymptotic2}---is violated. 

It is interesting to understand how this violation arises in this example and compare it to the heterotic string.
The explicit four-point open superstring amplitude is
\eq{
\begin{aligned}
 A =&\, K
\bigg[
{\rm tr}(T_1 T_2 T_3 T_4)\frac{\Gamma(-s)\Gamma(-t)}{\Gamma(1-s-t)}
 \\&\; + \left(\begin{array}{c} 1234\,{\rightarrow} \,1423 \\ s,t,u\,{\rightarrow }\,t,u,s \end{array}\right ) {+} \left(\begin{array}{c} 1234\,{\rightarrow} \,1342 \\ s,t,u\,{\rightarrow}\, u,s,t \end{array}\right )\bigg],
\end{aligned}
}{type1amp}
with $K\sim p^4$ encoding external polarizations.
Taking all polarizations transverse to the momenta, the Regge limit of Eq.~\eqref{type1amp} is a helicity factor  $\propto(\epsilon_1 \epsilon_4)(\epsilon_2 \epsilon_3)$ times
\eq{
\Gamma({-}t)\left[({-}s)^{1+t}(\mathbbm{P}_{\rm adj}^{(t)} \,{+}\, \mathbbm{X}) \,{-}\, ({-}u)^{1+t}(\mathbbm{P}_{\rm adj}^{(t)}\,{-}\,\mathbbm{X})\right],
}{reggesuperstring}
where $\mathbbm{P}_{\rm adj}^{(t)}=[{\rm tr}(T_1 T_2 T_3 T_4)-{\rm tr}(T_1 T_4 T_2 T_3)]/2$ and $\mathbbm{X}=[{\rm tr}(T_1 T_2 T_3 T_4)+{\rm tr}(T_1 T_4 T_2 T_3)]/2$. 
The Regge limit of $A^{(+)}$ is therefore $A^{(+)} \sim s^{1+t}  \mathbbm{X}$, scaling in the same way as $A^{(-)}\sim s^{1+t} \mathbbm{P}_{\rm adj}^{(t)}$. This equivalence is compatible with the $t$~pole being proportional to only $\mathbbm{P}_{\rm adj}^{(t)}$, and not to $ \mathbbm{X}$, because \Eq{reggesuperstring} reduces to $(s-u)\mathbbm{P}_{\rm adj}^{(t)}/t$ as $t\to 0^-$, thereby reproducing the gauge boson $t$-channel pole.

On the other hand, for closed strings the amplitude scales as $(su)^{1+t}$ in the Regge limit, with $s^2/t$ scaling in the forward limit signaling massless spin-two exchange. The leading color structure is fixed to $\mathbbm{1}$ by crossing symmetry, which forbids any antisymmetric color tensor that could cancel in the small-$t$ limit, and by the fact that the massless spin-two particle is a graviton, i.e., a singlet.

It would be interesting to attempt to construct an amplitude with a graviton that violates completeness or democracy. One possible approach is to take the ansatz from Ref.~\cite{Haring:2023zwu}, comprising a sum of stringy satellite terms with arbitrary coefficients, and dress it with color factors. However, it is easy to see that such an amplitude cannot be constructed from this ansatz. In the Regge limit, each satellite becomes $(-s)^{t+a}(-u)^{t+b}$ for some constants $a$ and $b$. One could dress the satellite terms so that the leading Regge behavior of the full amplitude is $(-s)^{t+a}(-u)^{t+b}(\mathbbm{1} +  \mathbbm{X}) + (-s)^{t+b}(-u)^{t+a}(\mathbbm{1} -  \mathbbm{X})$, which goes like $s^{2t+a+b}[(-1)^{a}(\mathbbm{1}+ \mathbbm{X})+(-1)^b(\mathbbm{1}- \mathbbm{X})]$. This expression is therefore proportional to either $\mathbbm{1}$ or $ \mathbbm{X}$, depending on the relative parity of $a$ and $b$. However, since this is the leading term, it must reproduce the graviton pole and therefore must be proportional to $\mathbbm{1}$, leaving representation completeness and democracy unmodified.

\end{document}